\pdfoutput=1
\documentclass[runningheads]{llncs}
\newif\ifTR
\TRfalse

\usepackage{fontawesome}

\usepackage{microtype}
\usepackage[frozencache,cachedir=.]{minted} % see https://github.com/gpoore/minted/issues/113#issuecomment-888045507
\usepackage{etex}
\usepackage{amsmath}
\usepackage{mathpartir}
\usepackage{stmaryrd}
\usepackage{xspace}
\usepackage{tikz}
\usepackage{mathbbol}
\DeclareMathSymbol{\fcmp}{\mathrel}{bbold}{\lq\;}
\usepackage{hyperref}
\usepackage{wasysym}
\usepackage{listings}
\usepackage{subcaption}
\usepackage{algorithm2e}

\newcommand{\eqdecl}[6]{
\centerline{$E=#1\qquad #2\rhd#3 = #4 \qquad #2\sqcup #3 = #5 \qquad I = #6$}
}

\begin{document}

\title{Error Localization for Sequential Effect Systems (Extended Version)}

\author{Colin S.\ Gordon\orcidID{0000-0002-9012-4490}
\and
Chaewon Yun
}
\authorrunning{Gordon and Yun}
\institute{
\email{\{csgordon,cy422\}@drexel.edu}\\
Drexel University, Philadelphia PA 19104, USA}
\maketitle

\begin{abstract}
    We describe a new concrete approach to giving predictable error locations for sequential (flow-sensitive) effect systems. 
    Prior implementations of sequential effect systems rely on either computing a bottom-up effect and comparing it to a declaration 
    (e.g., method annotation) or leaning on constraint-based type inference. 
    These approaches do not necessarily report program locations that precisely indicate where a program may ``go wrong'' at runtime.

    Instead of relying on constraint solving, we draw on the notion of a residual from literature on ordered algebraic structures. 
    Applying these to effect quantales (a large class of sequential effect systems) yields an implementation approach which accepts 
    exactly the same program as an original effect quantale, but for effect-incorrect programs is guaranteed to fail type-checking 
    with predictable error locations tied to evaluation order.
    We have implemented this idea in a generic effect system implementation framework for Java, and report on experiences applying 
    effect systems from the literature and novel effect systems to Java programs. We find that the reported error locations with 
    our technique are significantly closer to the program points that lead to failed effect checks.
\end{abstract}
\section{Introduction}

Effect systems are a well-established technique for extending a base type system that reasons about input and output shapes and available operations, to also statically reason about behaviors of code.  However, error reporting for effect systems has not been systematically studied. Existing implementations of effect systems report errors in one of two ways.

The classic approach is checking that each individual operation's effect is less than some bound~\cite{bocchino09,toro2015customizable,rytz12,gosling2014java,ecoop13} and reporting errors for any individual operation whose check fails. For example, this is how Java's checked exceptions are handled: every possibly-throwing expression's \mintinline{java}{throws} clause is checked against that of the enclosing method.  This yields highly precise error locations (e.g., reporting specific problematic method invocations or \mintinline{java}{throw} statements), but applies only for the (common) case of flow-insensitive effect systems.

The more general approach is to raise an error for whatever program point gave rise to a failing constraint during type inference~\cite{amtoft1999,nielson1993cml,bracker2015polymonadic_haskell,hicks2014polymonadic,orchard2014embedding,ivavskovic2020graded,ivavskovic2020data,talpin1992polymorphic}, which works for a wide variety of effect systems. However this leads to the well-known difficulty with localizing mistakes from type inference errors: the constraint which failed may be far away from an actual programmer mistake.  

Technically a third possibility is available, of comparing the computed effect of a method body against an annotated or assumed bound. This can work for any effect system, but we know of no implementations taking this approach, which would yield highly imprecise error messages (basically, ``this method has an effect error'').\footnote{Some implementations are designed for effect inference to always succeed, with a secondary analysis rejecting some effects outside the effect system~\cite{skalka2008types,skalka2020types}, and others with unavailable source do not provide enough detail to ascertain their effect checking algorithm~\cite{flanagan2003atomicity,boyapati02,flanagan2005type,flanagan2008types,sasturkar2005automated}.}

This is an unforunate state of affairs: less powerful effect systems have localized error reporting (the first approach), while more powerful flow-sensitive effect systems --- arguably in more dire need of precise error reporting --- are stuck with error reports that are unpredictable (approach 2) or maximally imprecise (approach 3).
This paper shows how to derive precise, predictable error reporting for flow-sensitive effect systems as well, by noticing that the first approach is in fact an error reporting optimization of the third: applied to the same effect system, they accept exactly the same programs, but while the third directly implements typical formalizations, the first in fact cleverly exploits algebraic properties of the third for more precise error reporting. By articulating and generalizing these properties, we obtain a new more precise error reporting mechanism for sequential~\cite{tate13} effect systems.

In the common case of an effect system where effects are partially ordered (or pre-ordered), while type-and-effect checking 
code with a known upper bound (such as a Java method with a throws clause), it is sufficient to check for each operation 
(e.g., method invocation or throw statement) whether the effect (e.g., the possibly-thrown checked exceptions) is less 
than the upper bound (e.g., \mintinline{java}{throws} clause).  If not, an error is reported for that operation.  This is a
deviation from how such effect systems are typically formalized, which is as a join semilattice, where formally the error 
would not occur until the least upper bound over \emph{all} subexpressions' effects was compared against the declared bound 
for the code.  Directly implementing this typical formalization is sound, but for a large method provides no direct clue as
to where the problematic code may be in the method.

This paper explores in detail why this optimization is valid, and uses that insight to generalize this precise error reporting to \emph{sequential} effect systems.  While the validity of this switch from joins in metatheory to local ordering checks in implementations is intuitively clear given basic properties of joins (namely, $\forall X,b\ldotp (\left(\bigsqcup X\right) \sqsubseteq b)\Leftrightarrow(\forall a\in X\ldotp a\sqsubseteq b)$), it is not obvious that there exists a corresponding transformation that can be applied to \emph{sequential} effect systems to move from global error checking to incremental error checking.  
Our contributions include:

\begin{itemize}
\item We explicitly identify and explain the common pattern of formalizing an effect system using operators that compute effects, while implementing the systems differently. We explain why this is valid.
\item We generalize this to arbitrary sequential effect systems characterized as effect quantales.
\item We describe an implementation of this approach for single-threaded Java programs, which is also the first implementation framework for sequential effect systems for Java.
\item We describe experiments implementing sequential effect systems in this framework and applying them to real Java programs, arguing that this theoretically grounded approach yields precise errors.
\end{itemize}

\section{Background}
Effect systems extend traditional type systems with information about side effects of program evaluation, which can be tailored to program behaviors of interest. Applications have included analyzing what regions of memory are accessed~\cite{lucassen88,gifford86,tofte1997region}; ensuring data race freedom~\cite{rccjava00,bocchino09,objtyrace99,Abadi2006}, deadlock freedom~\cite{suenaga2008type,safelocking99,tldi12,ivavskovic2020graded}, or other more targeted concurrency safety properties like safe use of GUI primitives~\cite{ecoop13,toro2015customizable,long2015intensional}; checking atomicity in concurrent programs~\cite{flanagan2003tldi,flanagan2003atomicity}; checking safety of dynamic software updates~\cite{neamtiu2008contextual}; checking communication properties~\cite{amtoft1999,nielson1993cml}; dataflow properties~\cite{ivavskovic2020data,bao21} or general safety properties of execution traces~\cite{skalka2008types,Skalka2008,Koskinen14LTR}.

Effect systems extend the typing judgement to include an additional component, the \emph{effect}, which is a syntactic description of an upper bound on an expression's behavior. The typical judgment form $\Gamma\vdash e : \tau \mid \chi$ is interpreted as meaning the under variable typing assumptions $\Gamma$, evaluating expression $e$ will produce a result of type $\tau$ (if execution terminates), exhibiting at most behaviors described by $\chi$.
Function types are also extended in effect systems to carry a \emph{latent} effect $\chi$, typically written superscript above a function arrow, as in $\tau\xrightarrow{\chi}\tau'$, indicating that $\chi$ is a bound on the function body's behavior, which the type rule for function application incorporates into the effect of function invocation.

Most often the (representations of) behaviors an effect system reasons about are assumed to form a join semilattice, and intuitively the effect of an expression is then the least upper bound (join) of the effects of all (executed) subexpressions. But this model of effect systems, while broad and including many useful and powerful effect systems, is incomplete.  Many of the effect systems of interest~\cite{suenaga2008type,tldi12,nielson1993cml,amtoft1999,skalka2008types,skalka2008types,Koskinen14LTR,flanagan2003tldi,flanagan2003atomicity,ivavskovic2020graded,neamtiu2008contextual,ivavskovic2020data,bao21} have additional structure because they track behaviors sensitive to evaluation order, and therefore have not only a partial ordering on effects to model a notion which behaviors subsume others, but also a notion of sequencing effects to track ordering of behaviour.  There is still some active debate as to what the appropriate common model of these \emph{sequential} effect systems should be. They are captured most generally by Tate's effectoids~\cite{tate13} (or equivalently, by the independently-proposed notion of polymonads~\cite{hicks2014polymonadic}), but these are typically acknowledged to be more general than most systems require. More pragmatic proposals include graded monads~\cite{katsumata14} and effect quantales~\cite{ecoop17,toplas21}, which differ primarily in what kinds of distributive laws are assumed (or not assumed) for how least-upper-bound and sequencing interact. Gordon~\cite{toplas21} gives a detailed survey of general models of sequential effect systems, and their relationships.

In the sequel, we work with sequential effect systems characterized by effect quantales, because their application to mainstream programming languages seems furthest-developed. For effect quantales there are general approaches to deriving treatments of loops~\cite{toplas21} as well as constructs derivable from tagged delimited continuations~\cite{ecoop20b} (e.g., exceptions, generators) from a basic effect quantale, while for related frameworks only single examples exist. Gordon~\cite{toplas21} also gives a survey of how a wide range of specific sequential effect systems from the literature are modeled by effect quantales.

\begin{definition}[Effect Quantale~\cite{toplas21}]
    An \emph{effect quantale} is a structure $Q=\langle E, \sqcup, \rhd, I\rangle$ composed of:
    \begin{itemize}
        \item A set of effects (behaviors) $E$
        \item A \emph{partial} join (least-upper-bound) $\sqcup:E\times E\rightharpoonup E$
        \item A \emph{partial} sequencing operator $\rhd:E\times E\rightharpoonup E$
        \item A unit element $I$
    \end{itemize}
    such that
    \begin{itemize}
        \item $\langle E,\rhd,I\rangle$ is a partial monoid with unit $I$ (i.e., $\rhd$ is an associative operator with left and right unit $I$)
        \item $\langle E,\sqcup\rangle$ is a partial join semilattice (i.e., $\sqcup$ is commutative, associative, and idempotent)
        \item $\rhd$ distributes over $\sqcup$ on both sides
        \begin{itemize}
            \item $x\rhd(y\sqcup z)=(x\rhd y)\sqcup(x\rhd z)$
            \item $(x\sqcup y)\rhd z=(x\rhd z)\sqcup(y\rhd z)$
        \end{itemize}
    \end{itemize}
\end{definition}
Note that when writing relations involving possibly-undefined expressions (e.g., since $x\sqcup y$ may be undefined), we consider two expressions equal if they both evaluate to the same element of $E$, or are both undefined.\footnote{Readers who tire of thinking about partiality can approximate this by imagining there is an additional distinguished error element, greater than all others, and preserved by all operators, representing undefined results, and which is invalid in later type rules. This is in fact, consistent with the original axiomatization~\cite{ecoop17}, though that definition complicates some metatheory.}

From the partial join we can derive a partial order on effects: $x\sqsubseteq y \Leftrightarrow x\sqcup y = y$. 
Again we must specify the meaning of $\sqsubseteq$ on possibly-undefined expressions: 
$x\sqsubseteq y$ is defined only if the join of $x$ and $y$ is defined.

Both $\rhd$ and $\sqcup$ are monotone in both arguments, in the sense that if $a\sqsubseteq b$ and $x\sqsubseteq y$, then $a\rhd x \sqsubseteq b \rhd y$ (when the right side is defined) and similarly for $\sqcup$.

Note that any join-semilattice-based effect system can be modeled in this system, using $\bot$ for identity, and reusing join for sequencing as well.

Gordon~\cite{toplas21} also describes properties of a partial unary iteration operation $-^*$ used to characterize loop effects, guaranteed to be:
\[\begin{array}{rl}
    \textrm{Extensive} & \forall x\ldotp x \sqsubseteq x^*\\
    \textrm{Idempotent} & \forall x\ldotp (x^*)^*=x^*\\
    \textrm{Monotone} & \forall x,y\ldotp x\sqsubseteq y \Rightarrow x^*\sqsubseteq y^*\\
    \textrm{Foldable} & \forall x\ldotp (x^*)\rhd(x^*)\sqsubseteq x^*\\
    \textrm{Possibly-Empty} & \forall x\ldotp I \sqsubseteq x^*
\end{array}\]
An optimally precise iteration operator can be derived for most effect quantales of interest~\cite{toplas21}: all finite effect quantales, and all effect quantales which have finite meets of elements greater than the unit element. This includes all prior specific effect systems considered in the survey section of that paper. 

\subsection{Implementing Effect Systems}
We concern ourselves with implementing effect system checking in a setting where the expected effect of every method in the
program is given, rather than inferred. This models a reasonable integration of effects into languages that require explicit
method signatures, consistent with a number of prior implementations of effect 
systems~\cite{toro2015customizable,ecoop13,gosling2014java,flanagan2003atomicity}. 
Experience has shown that while full inference has value, for many effect systems, a reasonable default (or local 
customization of defaults) is often sufficient to achieve modest annotation overhead. This also models scenarios where 
full inference has been employed, but manual annotations are required to refine undesirable inferences and force type 
checking to produce errors in a method that is intended to have a certain effect, but was inferred to have an incompatible 
effect.
We speak of methods because our prototype (Section \ref{sec:implementation}) targets Java, but the same principles would 
apply to procedural or functional languages.

\paragraph{Global Reporting}
Many formalizations of effect systems use a join-semilattice or effect-quantale-like formulation of the system, so an 
implementation can generally compute the effect of an expression bottom-up. The result is that for code with a fixed bound 
$\chi$ (say, the declared effect of a method), the implementation uses the $\sqcup$ and/or $\rhd$ operators to compute the 
body effect $\chi'$, and then once for each method checks that $\chi'\sqsubseteq\chi$. This is a direct implementation of 
common metatheory for effect systems, but as the only effect check occurs at the granularity of entire method or function 
bodies, there is only one possible error location for such a technique to report: the entire method or function body.
We are unaware of any concrete systems that explicitly acknowledge implementing this approach, though language in some 
papers is suggestive of such an approach (e.g., mentioning that ``inspection of a method'' with an error revealed a problem,
as opposed to indicating an error was flagged on a specific line of code). 

\paragraph{Precise Reporting for Commutative Effect Systems}
Most effect system implementations are limited to join-semilattice or partial order structures, which ignore program order.
Of those with available implementations, all we know of give precise error locations for code expressions which would lead 
to failing effect system checks by exploiting the trick mentioned in the introduction, that this class of system permits 
checking effects incrementally by checking if each subexpression's effect is less than the bound, since this is equivalent
 to computing the join of subexpression effects and comparing to the bound. This is true of all available implementations we are aware of, including Java's checked exceptions~\cite{gosling2014java}, the modern implementation of Gordon et al.'s UI effects~\cite{ecoop13} in the Checker Framework~\cite{PapiACPE2008,DietlDEMS2011}, Toro and Tanter's framework~\cite{toro2015customizable} for gradual effects~\cite{BanadosSchwerter2014gradual}, Rytz et al.'s work for Scala~\cite{rytz12},
Deterministic Parallel Java~\cite{bocchino09}, and others.
As mentioned earlier, this trick does not work for general sequential effect systems.

\paragraph{Constraint-Based Reporting}
Most prior implementations of sequential effect systems --— and \emph{all} prior implementations \emph{frameworks} for 
sequential effect systems~\cite{hicks2014polymonadic,bracker2015polymonadic_haskell,bracker2016supermonads,bracker2018supermonads,orchard2014embedding}
--- use type inference to infer effects.  This results in the standard trade-offs for global constraint-based type 
inference: types (and effects) are inferred with low developer effort when possible, but errors can be cryptic, and 
implicate program locations unrelated to the error.  In particular, these implementations tend to generate subtyping
(and subeffecting) constraints from the program structure, which are then solved incrementally by a fixpoint solver.
Errors are reported at the location corresponding to the first constraint which is found to be inconsistent.
However, that constraint may be totally unrelated to any problematic statement.
Consider the brief JavaScript program\\
\centerline{\mintinline{javascript}{var x = 3; var y = x; requiresString(y)}}
It is possible for a constraint solver to flag any of the three statements as a type error, assuming the invoked method is
typed as requiring a string input. Flagging the first or third lines is reasonable, as they are the sources of the contradiction.
However, solvers are permitted to report the middle line as erronneous as well (for storing a number into a string-containing
variable), which is not terrible in this case, but becomes problematic with larger blocks of code.
This has inspired a wealth of work on various techniques to reduce or partially compensate for (but not eliminate) this 
unpredictability~\cite{pavlinovic2014finding,pavlinovic2015practical,loncaric2016practical,hassan2018maxsmt,lerner2007searching,chen2014counter}.
In principle such work is applicable to existing approaches to inferring effects in sequential effect 
systems~\cite{flanagan2008types,orchard2014embedding,nielson1997communication}, but unpredictability would remain.

\section{Local Errors for Sequential Effect Systems}
We would like to obtain precise error reporting for sequential effect systems in general. Because effect 
quantales subsume the join semilattice model of effect systems~\cite{toplas21}, we can hope to draw some 
inspiration from the corresponding optimization on traditional commutative effect systems: that optimization 
should be a special case of a general solution.

Let us fix an expression $e$ whose effect we would like to ensure is less than $\chi$.  
Let us assume a set $\{\chi_i\mid i\in\mathsf{Subterms}(e)\}$ where $\chi_i$ is the static effect of subterm 
$i$ from a bottom-up effect synthesis.  For now, we will assume all such effects are defined (i.e., that the 
bottom-up synthesis of effects never results in undefined effects).  For the case where effects form a 
join-semilattice, and all bottom-up computation is joins (no other operators play a role), we can formally 
relate the global and precise implementations of join-semilattice effect systems by observing

$$\bigsqcup\{\chi_i\mid i\in\mathsf{Subterms}(e)\}\sqsubseteq \chi \Leftrightarrow\forall i\in\mathsf{Subterms}(e)\ldotp \chi_i\sqsubseteq\chi$$

as suggested in the introduction. The left side of the iff expresses the global view that the join of all subexpression effects must be bounded by $\chi$.  The right side expresses the local view that each individual subexpression's effect must be less than the bound $\chi$.  (This formulation suggests some redundant checks; we return to this later.) One way to express the intuition behind this formula is that performing all of the local checks corresponds to ensuring that each individual $\chi_i$ can be further combined with some other effects (here, by join), and the result will still be bounded by $\chi$.  Conversely, if there exists some $j\in\mathsf{Subterms}(e)$ such that $\chi_j\not\sqsubseteq\chi$, then no combination with other effects can yield something satisfying the bound $\chi$.

We dub this informal characterization as the notion of \emph{completing} an effect. An effect $\chi_i$ \emph{can be completed to} $\chi$ if there exists some effect $\chi'$ such that the combination of $\chi_i$ and $\chi'$ is $\sqsubseteq\chi$. The general intuition is that if $\chi_i$ can be completed to $\chi$, it is possible to ``add more behaviors'' to $\chi_i$ and obtain an effect less than $\chi$, in the sense that it is possible to extend a program with effect $\chi_i$ with additional behaviors such that the overall effect is less than $\chi$. We formalize this later in this section, but to do so we must recall and customize a bit of relevant math.

\subsection{Residuals}

The literature on ordered semigroups~\cite{birkhoff} (particularly on non-commutative substructural logics~\cite{lambek1958mathematics,galatos2007residuated}) contains many applications of the notion of a \emph{residual}~\cite{ward1939residuated,dilworth1939non}:

\begin{definition}[(Right) Residual]
\label{def:right_residual}
A (right) \emph{residual} operation on an ordered monoid M is a binary operation $-\setminus-:M\times M\rightarrow M$ such that for any x, y, and z, $x\le y\setminus z \Leftrightarrow y\cdot x\le z$
\end{definition}

That is, the right residual $y\setminus z$ of $z$ by $y$ (also read as $y$ under $z$) is
an element of $M$ such that, when sequenced \emph{to the right} of $y$, yields an element no greater than $z$
(but definitely ordered $\le z$).

We can adapt this for effect quantales as well:

\begin{definition}[(Right) Quantale Residual]
    \label{def:total-quantale-residual}
A (right) residual operation on an effect quantale Q is a partial binary operation $-\setminus-:Q\times Q\rightharpoonup Q$ such that for any x, y, and z, $x\sqsubseteq y\setminus z \Leftrightarrow y\rhd x\sqsubseteq z$. A \emph{(right) residuated effect quantale} is an effect quantale with a specified choice of (right) residual operation.
\end{definition}

Consider the case of type-checking $e_1;e_2$, where $\Gamma\vdash e_1 : \chi_1$ and $\Gamma\vdash e_2 : \chi_2$, and ensuring
that the effect of the sequential composition of these expressions — $\chi_1\rhd\chi_2$ — is bounded by $\chi$.  A residual on
effects can tell us if this is \emph{possible} based on analyzing \emph{only} $e_1$,
in some cases rejecting programs before even analyzing $e_2$.

If $\chi_1\setminus \chi$ is undefined, then there \emph{does not exist} $\chi'$ such that $\chi_1\rhd\chi'\sqsubseteq\chi$. If there were some such $\chi'$, then by the definition of the residual operation, $\chi'\sqsubseteq \chi_1\setminus\chi$, which would imply the residual was defined. Thus an implementation could eagerly return an error after synthesizing the effect $\chi_1$ for $e_1$, and determining the residual was undefined.
This is a subtle point about the definition above: it states not only properties of the residual when it is defined, but also requires it to be defined in certain cases.

Many effect quantales have a right residual in the sense of Definition \ref{def:right_residual}, 
but not all do, and since effect systems are primarily concerned with sound bounds on behavior rather than exact characterizations 
of behavior, we actually require only a slightly weaker variant of residual:
\begin{definition}[Weak (Right) Quantale Residual]
\label{def:weak_right_residual}
A weak (right) residual operation on an effect quantale Q is a partial binary operation $-\setminus-:Q\times Q\rightharpoonup Q$ such that for any x, y, and z:
\begin{itemize}
    \item Residual bounding: $x\sqsubseteq y\setminus z \Rightarrow y\rhd x\sqsubseteq z$
    \item Residual existence: $y\rhd x\sqsubseteq z \Rightarrow \exists r\ldotp r=y\setminus z$
    \item Self-residuation: $\exists r\ldotp z\setminus z = r$
    \item Unit residuation: $I\setminus z = z$
\end{itemize}
A \emph{(right) residuated effect quantale} is an effect quantale with a specified choice of weak (right) residual operation.
\end{definition}

This is the notion of residual we work with, and this weakening is necessary to capture aspects of non-local control flow~\cite{ecoop20b}.
Every residual operation in the remainder of this paper is a weak residual, though for brevity we simply refer to them as residuals. The \emph{total} residual (Definition \ref{def:total-quantale-residual}) implies the axioms of the weak residual, so in some cases we present a total residual as a weak residual.

We take this weak right residual to be our formal notion of completion: an effect $\chi$ can be completed to $\chi'$ if the weak residual $\chi\setminus\chi'$ is defined.

It is worth noting that the literature also contains a definition of left residual, which we could use to similarly issue an eager warning given only $\chi$ and $\chi_2$. However, notice that the right residual corresponds to type-checking traversals proceeding in the standard left-to-right evaluation order standard in (most) languages using call-by-value reduction.  Because most other analysis tools, and in practice most developer investigation of program-order dependent behaviors proceeds in tandem with evaluation order, we focus solely on the right residual. However, all results in the rest of the paper can be dualized to the left residual.  Because we focus exclusively on the right residual, for the rest of the paper we will drop the qualifier ``right'' and simply refer to unqualified residuals.

This seems a promising approach, but detailing it fully requires also connecting our notion of completion to the way effects are actually combined during type-checking (e.g., most programs are not basic blocks of primitive actions).  Before doing so, we build further intuition by describing the residual operations for a few existing effect quantales in the literature, based on Gordon's formalization~\cite{toplas21}.

\subsection{Residual Examples}
This section gives several examples of residuated effect quantales, to show that many existing effect systems already naturally satisfy the requirements of our weak residual,
so while it is not mathematically the case that all effect quantales have a weak residual, it appears known effect systems typically do.
Appendix \ref{apdx:more} contains additional examples.

\subsubsection{Traditional Commutative Effects}

In the case where the effect quantale is simply a (partial) join semilattice (so $\rhd=\sqcup$), the residual is simply:

$$X\setminus Y = Y ~~\textrm{when}~~ X \sqsubseteq Y$$

Thus the residual is exactly the local subeffect comparison performed by local implementations of effect checking. 

\subsubsection{Formal Languages and Quotients Thereof}
\label{sec:formal_lang}
As formal languages can model sets of acceptable behaviors (and are often used for this purpose, most frequently
in automata-theoretic model checking), it is worth considering formal languages as effects. Indeed,
languages of finite words over a finite alphabet form an effect quantale:
Here we recall a distillation of a number of general behavioral trace effect systems~\cite{skalka2008types,Skalka2008,Koskinen14LTR} given by Gordon~\cite{toplas21}:
\begin{definition}[Finite Trace Effects]
    \label{def:fintraceeq}
    Effects tracking sets of finite event traces, for events in an alphabet $\Sigma$, form an effect quantale:\\
    \eqdecl{\mathcal{P}(\Sigma^*)\setminus\emptyset}{X}{Y}{X\cdot Y}{X\cup Y}{\{\epsilon\}}
    Where sequencing is pairwise concatenation of sets, $X\cdot Y=\{xy \mid x\in X\land y\in Y\}$.
\end{definition}
We write this effect quantale for a particular alphabet $\Sigma$ as $\mathsf{FinTrace}(\Sigma)$.

Intuitively, the residual should be defined whenever the dividend is some kind of prefix of all behaviors in the 
numerator. The formalization is more subtle, but captures this intuition:
\[
    X\setminus Y = \{ w\in\Sigma^* \mid \forall x\in X\ldotp x\cdot w \in Y\}~\textrm{when non-empty}
\]
That is, the residual $X\setminus Y$ is the set of words which, when sequenced after $X$, will produce a subset of $Y$.

Note that this is \emph{not} the quotient of formal languages, which uses the same notation with a different meaning.
The (right) \emph{quotient} of X under Y $X\setminus_q Y$ (using the subscript $q$ to distinguish the quotient from the
residual) is $\{w\in\Sigma^*\mid\exists x\in X\ldotp x\cdot w\in Y\}$. 
Sequencing this after the set $X$ yields $\{xw\mid x\in X\land \exists x'\in X\ldotp x'\cdot w\in Y\}$, which may be larger 
than $Y$.

The above (full) residual is in fact an operation that exists in Action Logic, a cousin of Kleene Algebra.
Pratt~\cite{pratt1990action} notes that the two are related; in our notation, $X\setminus Y=(X\setminus_q(Y^{-1}))^{-1}$.
Since regular languages are closed under complement and quotient, this means they are also closed under
residuation, and therefore there is a sub-effect-quantale $\mathsf{Reg}(\Sigma)\subseteq\mathsf{FinTrace}(\Sigma)$
which \emph{also} has weak residuals, which can be computed using operations on finite automata (and mapped back to regular expressions
for error reporting).

The induced iteration operator on this effect quantale corresponds to the Kleene star, so in later examples we sometimes use regular expression syntax for writing these effects.  
Note, however, that while these effect quantale operations correspond conveniently to regular expressions, the 
effect quantale itself is not limited to regular languages: concatenation, union, and Kleene iteration are well-defined 
operations on \emph{any} formal languages, including context-free, context-sensitive, or recursively enumerable languages. 
Regular languages are restricted to be built from \emph{only} these operations and singleton sets, but no such assumption 
is present here.

\subsection{Atomicity}
\label{sec:atomicity_bg}
\begin{figure}[t]
\vspace{-0.5cm}
\begin{subfigure}[b]{0.3\textwidth}
\begin{center}
\begin{minipage}{0.3\textwidth}
\begin{tikzpicture}
\node(Top){$\top$};
\node(A)[below of=Top]{$A$};
\node(R)[below right of=A]{$R$};
\node(L)[below left of=A]{$L$};
\node(B)[below right of=L]{$B$};
\draw(B)--(L);
\draw(B)--(R);
\draw(R)--(A);
\draw(L)--(A);
\draw(A)--(Top);
\end{tikzpicture}
\\
$
\begin{array}{|c|ccccc|}
\hline
\rhd & B & L & R & A & \top\\
\hline
B & B & L & R & A & \top\\
R & R & A & R & A & \top\\
L & L & L & \top & \top & \top\\
A & A & A & \top & \top & \top\\
\top & \top & \top & \top & \top & \top\\
\hline
\end{array}
$
\end{minipage}
\end{center}
\caption{Atomicity effects~\cite{flanagan2003tldi}}
\label{fig:atomicity_lattice}
\end{subfigure}
~~~~
\begin{subfigure}[b]{0.60\textwidth}
\begin{center}
\begin{tikzpicture}[node distance=1.5cm]
\node(neutral){\fbox{$\varepsilon$}};
\node(critical)[above left of=neutral]{\fbox{$\mathsf{critical}$}};
\node(locking)[left of=critical]{\fbox{$\mathsf{locking}$}\hspace{2em}};
\node(entrant)[above right of=neutral]{\fbox{$\mathsf{entrant}$}};
\node(unlocking)[right of=entrant]{${}\qquad{}$\fbox{$\mathsf{unlocking}$}};
\draw(neutral)--(critical);
\draw(neutral)--(entrant);
\end{tikzpicture}
\\
$
\begin{array}{|c|ccccc|}
\hline
\rhd & \mathsf{locking} & \mathsf{unlocking} & \mathsf{critical} & \mathsf{entrant} & \varepsilon \\
\hline
\mathsf{locking} & - & \mathsf{entrant} & \mathsf{locking} & - & \mathsf{locking} \\
\mathsf{unlocking} & \mathsf{critical} & - & - & \mathsf{unlocking} & \mathsf{unlocking} \\
\mathsf{critical} & - & \mathsf{unlocking} & \mathsf{critical} & - & \mathsf{critical} \\
\mathsf{entrant} & \mathsf{locking} & - & - & \mathsf{entrant} & \mathsf{entrant} \\
\varepsilon & \mathsf{locking} & \mathsf{unlocking} & \mathsf{critical} & \mathsf{entrant} & \varepsilon \\
\hline
\end{array}
$
\end{center}
\caption{Critical section effects~\cite{tate13}.}
\label{fig:crit-effects}
\end{subfigure}
\caption{
    Lattices and sequencing for atomicity and critical section (reentrancy) effects.
    $-$ represents an undefined result for sequential composition.
}
\label{fig:lattices}
\end{figure}
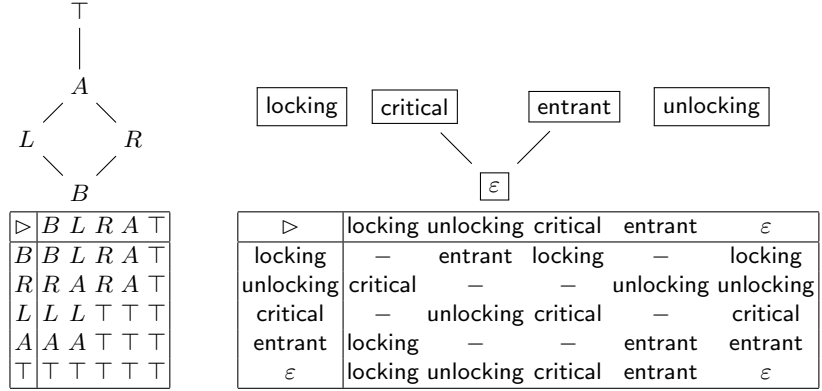
Flanagan and Qadeer~\cite{flanagan2003tldi,flanagan2003atomicity} proposed well-known approaches to capturing atomicity as
effects. Their original proposal turns out to be a particular finite effect quantale~\cite{ecoop17,toplas21},
whose join semilattice and sequential composition are shown in Figure \ref{fig:atomicity_lattice}\footnote{
    They subsequently proposed an extension to conditional atomicity~\cite{flanagan2003atomicity},
        which is also an effect quantale when combined with a data race freedom quantale.
}]
(in the bottom table, the effect in row $i$ sequenced via $\rhd$ with the effect in column $j$ is equal
to the effect in cell $i,j$ of the table).
The key idea is to adopt Lipton's theory of reduction~\cite{lipton75} to label each expression with an effect
capturing how it commutes with shared-memory operations in other threads: $B$ for both directions (e.g.,
thread-local actions), $L$ for left
(i.e., earlier, such as lock releases), right ($R$, later, such as lock acquisitions), atomic $A$ 
(does not commute, including atomic hardware operations and already-proven-atomic critical sections),
or compound $\top$ (interleaves in non-trivial ways with other threads).

Sequential composition captures Lipton's idea that any sequence of $R$ or $B$ actions, followed by at most one
atomic $A$ action, then any sequence of $L$ or $B$ actions, can be grouped together as if the whole sequence
occurred atomically from the perspective of another thread.
We omit iteration for brevity, but the original definition of $(-)^*$ coincides with the results of a general
construction on finite effect quantales as well~\cite{toplas21}.

We can define a (total) residual $\chi\setminus\chi'$ according to the classic mathematical definition:
$\chi\setminus\chi' = \bigsqcup \{\chi'' \mid \chi\rhd\chi''\sqsubseteq\chi'\}$,
which is well-defined because the join semilattice is complete. So for example, $L\setminus A$ is the
greatest effect which, when sequenced \emph{after} $L$, yields a result less than $A$ --- which in this
case works out to be $L$ itself.
 
\subsection{Reentrancy}
\label{sec:reentrancy_bg}
Tate~\cite{tate13} developed a maximally-general framework for sequential effect systems, and his running
example was the system given in Figure \ref{fig:crit-effects}, which has partial joins and sequencing.
(He did not describe an iteration operator, but one can be derived from the general construction of iteration
operations on finite effect quantals~\cite{toplas21}.)
This is an effect system motivated by tracking critical sections for a single global resource lock
which does \emph{not} permit recursive acquisition.
This turns out to \emph{also} be a natural effect system for tracking non-reentrant code:
the start of a non-reentrant operation can be given the effect \textsf{locking}.
Notice that $\mathsf{locking}\rhd\mathsf{locking}$ is undefined, so attempting to
reenter an operation that is non-reentrant will not type-check.
This covers non-reentrant locks (as in the original example, and in implementations such as Java's
\lstinline|StampedLock|), but also database APIs that do not
support nested transactions (starting and finishing transactions),
or the evaluation API for Java's \texttt{XPath} expressions (\lstinline|XPathExpression.evaluate(...)|).

\subsection{Connecting Residuals to Type Checking}

\newcommand{\extend}[1]{\textcolor{blue}{#1}\xspace}

\begin{figure}[t]
    \begin{mathpar}
\begin{array}{lrcl}
\textsc{Effects} & \chi & \in & Q~\textrm{(effect quantale)}\\
\textsc{Types} & \tau & ::= & \mathsf{bool} \mid \mathsf{unit} \mid \tau\xrightarrow{\chi}\tau\\
\textsc{Expressions} & e & ::= & x \mid c \mid \lambda^{\chi}x:\tau\ldotp e \mid e@e \mid \mathsf{if}~e~\textrm{then}~e~\textrm{else}~e \mid \mathsf{while}~e~e\\
\end{array}
\and
\fbox{$\extend{\chi\mapsto\chi\parallel{}}\Gamma\vdash e : \tau \mid \chi$}
\and
\inferrule*[left=T-Var]{\Gamma(x)=\tau}{\extend{\chi_0\mapsto\chi_m\parallel{}}\Gamma\vdash x : \tau \mid I }
\and
\inferrule[T-Const]{ }{\extend{\chi_0\mapsto\chi_m\parallel{}}\Gamma\vdash c : \tau_c \mid I }
\and
\inferrule*[left=T-Lambda]{
    \extend{I\mapsto\chi\parallel{}}\Gamma,x:\tau\vdash e : \tau'\mid \chi' \\ \chi'\sqsubseteq\chi
}{
    \extend{\chi_0\mapsto\chi_m\parallel{}}\Gamma\vdash \lambda^\chi x:\tau\ldotp e: \tau\overset{\chi}{\rightarrow}\tau' \mid I
}
\and
\inferrule*[left=T-App]{
    \extend{\chi_0\mapsto\chi_m\parallel{}}\Gamma\vdash e_1 : \tau' \overset{\chi_l}{\rightarrow}\tau\mid \chi_1\\
    \extend{\chi_0\rhd\chi_1\mapsto\chi_m\parallel{}}\Gamma\vdash e_2:  \tau\mid\chi_2\\
    \extend{\chi_l\setminus(\chi_2\setminus(\chi_1\setminus(\chi_0\setminus\chi_m)))~\mathsf{defined}}
}{
    \extend{\chi_0\mapsto\chi_m\parallel{}}\Gamma\vdash e_1@e_2 : \tau\mid\chi_1\rhd\chi_2\rhd\chi_l
}
\and
\inferrule*[left=T-If]{
    \extend{\chi_0\mapsto\chi_m\parallel{}}\Gamma\vdash e_c : \mathsf{bool} \mid \chi_c \\
    \extend{\chi_0\rhd\chi_c\mapsto\chi_m\parallel{}}\Gamma\vdash e_t : \tau \mid \chi_t \\
    \extend{\chi_0\rhd\chi_c\mapsto\chi_m\parallel{}}\Gamma\vdash e_f : \tau \mid \chi_f \\
    \extend{(\chi_t\sqcup\chi_f)\setminus(\chi_c\setminus(\chi_0\setminus\chi_m))~\mathsf{defined}}
}{
    \extend{\chi_0\rhd\chi_m\parallel{}}\Gamma\vdash \mathsf{if}~e_c~\mathsf{then}~e_t~\mathsf{else}~e_f : \tau \mid \chi_c\rhd(\chi_t\sqcup\chi_f)
}
\and
\inferrule[T-While]{
    \extend{\chi_0\mapsto\chi_m\parallel{}}\Gamma\vdash e_c : \mathsf{bool} \mid \chi_c \\
    \extend{\chi_0\rhd\chi_c\mapsto\chi_m\parallel{}}\Gamma\vdash e_b : \tau \mid \chi_b \\
    \extend{(\chi_b\rhd\chi_c)^*\setminus(\chi_c\setminus(\chi_0\setminus\chi_m))~\mathsf{defined}}
}{
    \extend{\chi_0\mapsto\chi_m\parallel{}}\Gamma\vdash\mathsf{while}~e_c~e_b : \mathsf{unit}\mid \chi_c\rhd(\chi_b\rhd\chi_c)^*
}
    \end{mathpar}
    \caption{Type rules with and without residual checks.}
    \label{fig:typing}
\end{figure}

Figure \ref{fig:typing} defines two typing judgments.  $\Gamma\vdash e : \tau \mid \chi$ is a standard judgment form for effect systems, interpreted as ``under variable typing assumptions $\Gamma$, expression $e$ has type $\tau$ with effect $\chi$.''
This judgment is readable in Figure \ref{fig:typing} by ignoring the extensions in \textcolor{blue}{blue}, and corresponds to a subset of the type rules Gordon~\cite{toplas21,ecoop17} proved sound for a wide array of possible primitives and state models. In short, the judgment types expressions, using the effect quantale operators to synthesize the effect of the expression, capturing evaluation ordering with $\rhd$ and alternative paths (e.g., in \textsc{T-If}) with $\sqcup$.  We call this judgment the \emph{standard judgment}.

Including the text in \textcolor{blue}{blue}, Figure \ref{fig:typing} defines an additional judgment $\chi_0\mapsto\chi_m\parallel\Gamma\vdash e : \tau \mid \chi$, interpreted as
``under typing assumptions $\Gamma$ expression $e$ has type $\tau$ and effect $\chi$, and moreover if $e$ is executed after effect $\chi_0$, it is still possible for the result to have effect less than $\chi_m$.'' Later we formalize this  interpretation.
This extended judgment form performs additional checks (also in \textcolor{blue}{blue}).
These additional checks do not reject (or accept) any additional programs, but restrict how, when, and why programs are rejected.

\begin{lemma}[Conservative Extension \faGear]
    \label{lem:extended_implies_standard}
    If $\chi_0\mapsto\chi_m\parallel\Gamma\vdash e : \tau \mid \chi$, then $\Gamma\vdash e : \tau \mid \chi$.
\end{lemma}
\ifTR
\begin{proof}
    By induction over the extended judgment. Intuitively, ignoring the additional checks leaves the standard judgment.
\end{proof}
\else
\fi
This lemma, and others marked with a \faGear{} icon, have been mechanically checked in the \textsc{Coq} proof assistant.

Critically, this result implies than any soundness results holding for the standard judgment are inherited by the extended judgment. Gordon~\cite{toplas21} gives generic type safety results for a configurable framework of sequential effect systems, parameterized by primitives and choices of states. Since our standard judgment is an instantiation of that framework (for a specific choice of constants), our extended judgment is also type-safe for appropriate choices of state and primitive semantics. We do not give further consideration to type-safety in this paper, as Gordon's framework can be instantiated to yield type-safety results for all of our examples effect systems (in some cases, demonstrated in that work~\cite{toplas21}).

Proving the other direction of the correspondence relies on weakened versions of standard residual properties:
\begin{lemma}[Residual Sequencing \faGear]
\label{lem:resid_seq}
For any effects $x,y\in Q$ for a residuated effect quantale $Q$, if $x\setminus y$ is defined, then $x\rhd(x\setminus y) \sqsubseteq y$. 
\end{lemma}
\ifTR
\begin{proof}
    This follows from residual bounding since $(x\setminus y)\sqsubseteq(x\setminus y)$.
\end{proof}
\fi
\begin{lemma}[Residual Shifting \faGear]
\label{lem:residual_shifting}
For any effects $x,y,z\in Q$ for a residuated effect quantale $Q$, $(x\rhd y)\setminus z$ is defined if and only if $y\setminus(x\setminus z)$ is defined. 
\end{lemma}
\ifTR
\begin{proof}
To prove the forward direction $y\setminus(x\setminus z)$:
    Trivially, we know $y\setminus(x\setminus z)\sqsubseteq y\setminus(x\setminus z)$ (recall this does not require the expression to be defined). This is enough to conclude via residual bounding that $y\rhd(y\setminus(x\setminus z))\sqsubseteq x\sqsubseteq z$. This ordering can be used with residual bounding to conclude that the residual of the right hand side with $y$ --- our goal --- is defined.

To prove the reverse, residual existence implies the residual is defined if we can exhibit some $q$ such that $(x\rhd y) \rhd q \sqsubseteq z$. Letting $q$ be $y\setminus(x\setminus z)$, we can prove this via transitivity, stepping through:
\\\centerline{$(x\rhd y)\rhd (y\setminus (x\setminus z)) \sqsubseteq x\rhd (x\setminus z) \sqsubseteq z$}
The left ordering follows from associativity and residual sequencing (Lemma \ref{lem:resid_seq}). The right follows again from residual sequencing.
\end{proof}
\fi

\begin{lemma}[Antitone Residuation \faGear]
    \label{lem:antitone_resid}
    If $x\sqsubseteq y$ and $y\setminus z$ is defined, then $x\setminus z$ is defined.
\end{lemma}
\ifTR
\begin{proof}
    If we can exhibit a $q$ such that $x\rhd q\sqsubseteq z$, this follows via residual existence.
    Let $r=y\setminus z$. Then $r\sqsubseteq y\setminus z$, and by residual bounding, $y \rhd r \sqsubseteq z$. By monotonicity of $\rhd$, $x\rhd r\sqsubseteq y\rhd r$, so transitively $x\rhd r\sqsubseteq z$.
    From this and residual existence we can conclude $x\setminus z$ is defined.
\end{proof}
\fi
These are enough to prove that the standard typing judgment implies the extended judgment holds, for reasonable choices of $\chi_0$ and $\chi_m$. By reasonable, we mean that it is possible to ``reach'' $\chi_m$ by running code with effect $\chi_0$, then code with the effect the standard judgment assigns, followed by some additional effect, without exceeding a total upper bound of $\chi_m$.
\begin{lemma}[Liberal Extension \faGear]
    \label{lem:extension}
    If $\Gamma\vdash e : \tau \mid \chi$, then for any $\chi_0$ and $\chi_m$ such that $\chi\setminus(\chi_0\setminus\chi_m)$ is defined, $\chi_0\mapsto\chi_m\parallel\Gamma\vdash e : \tau \mid \chi$.
\end{lemma}

\paragraph{Loop Unrolling}
While not required for the proof above, readers may wonder if the desired residual is provable for finite loop unrollings, i.e., if $(\chi_b\rhd\chi_c)^n\setminus(\chi_c\setminus(\chi_0\setminus\chi_m))$ is defined for all naturals $n$. Indeed it is: the properties of iteration imply $(\chi_b\rhd\chi_c)^n\sqsubseteq(\chi_b\rhd\chi_c)^*$, so this follows from Lemma \ref{lem:antitone_resid}.

\paragraph{Completability}
These results highlight that the extended judgment accepts exactly the same programs as the standard judgment, but ensures our informally-stated requirement that the extended judgment checks that given a prefix effect and target bound, the program has a valid effect in that usage context:
\begin{theorem}[Completability \faGear]
    \label{thm:completability}
    When $\chi_0\setminus\chi_m$ is defined and $\chi_0\mapsto\chi_m\parallel\Gamma\vdash e : \tau \mid \chi$, then 
    $\chi\setminus(\chi_0\setminus\chi_m)$ is defined.
\end{theorem}
\ifTR
\begin{proof}
    By induction on the extended judgment. \textsc{T-Var}, \textsc{T-Const}, and \textsc{T-Lambda} satisfy the result trivially (residuals by unit always exist). \textsc{T-App}, \textsc{T-If}, and \textsc{T-While} follow by applying the defining residual property to the residual check in the rule.
\end{proof}
\fi
This result highlights a subtlety of the nested residual checks in several rules. While at a glance it may appear to lump many checks into one since the residuals checked for existence are larger nested residuals, these checks actually verify only one residual definition beyond what is guaranteed by the subexpressions' typing results.
Consider as an example \textsc{T-App}. If the antecedent subexpression typings hold, then by Completability $(\chi_1\setminus(\chi_0\setminus\chi_m))$ is defined and $\chi_2 \setminus (\chi_1\setminus(\chi_0\setminus\chi_m))$ is defined. So \textsc{T-App} is truly checking only the residual with respect to the latent effect of the invoked function.
Likewise, \textsc{T-If} and \textsc{T-While} are checking only the residual with respect to the respective constructs, with residuals for the subexpression effects already guaranteed by the antecedents. Later this allows us to ensure that algorithmically, only one residual check is required per source construct, and that if it would fail due to a subexpression rather than the construct being checked, that error would have already been reported when checking the subexpressions individually.

The combination of Theorem \ref{thm:completability} and Lemma \ref{lem:extension} guarantees that if an expression
is well-typed under standard typing, but not under extended typing for a given $\chi_0$ and $\chi_m$ with $\chi_0\setminus\chi_m$
defined, it is because the subexpressions are arranged in a way that is incompatible with the 
effect context $\chi_0\mapsto\chi_m$. And because the residual checks in each rule fail (the residual becomes undefined)
as soon as any execution prefix's residual is undefined, if the extended judgment rejects a term accepted by
the standard judgment, the specific (nested) residual check pinpoints the inadmissible execution prefix for that
context.

\subsection{Limitations}
For the residual to localize errors effectively, it must be the case that some residuals are undefined, since if all 
residuals exist, then there is no basis for issuing the early errors we seek. A trivial example is the effect quantale 
given by a group with only the reflexive ordering.  In such an effect quantale, it is always possible to start with a 
prefix effect $X$ and reach a final effect $Y$ by sequencing $X^{-1}Y$ after $X$, so no residual check will fail --- 
only the final subeffect check in \textsc{T-Lam}.
While such effect quantales exist mathematically, we are unaware of any such quantales actually being used as
effect systems, as conceptually, they are at odds with the typical goal of an effect system, which is to build a 
sound summary of (selected) program behavior and reject certain subsets.

Kleene Algebras~\cite{kozen1997kleene} are iterable effect quantales~\cite{toplas21} which are typically used in program analysis,
and typically include a least element in lattice order which corresponds to ``no behavior'' (e.g., an empty set
or empty relation). Since most programs are expected to have some kind of behavior,
one could modify a Kleene Algebra to be partial (i.e., removing the least element from its domain) and obtain
an effect quantale with meaningful notions of residuals. However, even then there may be ``too many'' residuals
to be useful. For example, in a Kleene Algebra of binary relations (pre- and post-conditions) or 
of transition functions (e.g., modeling abstract interpretation as a Kleene Algebra~\cite{kot2004second}),
the domain is large enough to admit arbitrary transitions: for any prefix behavior with precondition 
$P$ as a set of states and ultimate goal postcondition $Q$ as a set of states, the relation $P\times Q$
is non-empty as long as both $P$ and $Q$ are non-empty. So applying our technique to domains like these
notions of extensional correctness would require refining the set of transition functions or relations considered,
in order to make the residual undefined when no action contained in the actual program could lead to a desired execution.
This is less problematic for intensional specifications such as the language-theoretic approaches, because
as soon as some execution path commits to a behavior prefix not in the target specification, the residual is undefined.

\section{Algorithms}
To turn our insights into a tool, we require a type-checking algorithm.
Figure \ref{fig:algo} shows the core pieces of a type-checking algorithm that accepts exactly the programs well-typed under the extended typing judgment of Figure \ref{fig:typing}.\footnote{The figure uses Java 17's extended switch statements, plus a few notational liberties for operations on effect quantale elements.}
We assume that \mintinline{java}{earlycheck} accepts the contextual and goal effects, type environment, and expression as inputs, and returns a result (a supertype of successful typing results, unbound variable errors, type errors, and effect errors).
\begin{figure}[t!]
\begin{minted}[escapeinside=//,mathescape=true,fontsize=\small]{Java}
Result earlycheck(Q /$\chi_0$/, Q /$\chi_m$/, env /$\Gamma$/, expr e) {
  switch (e) {
    case Var x:
      Type t = lookup(/$\Gamma$/, x);
      return (t != null) ? new Result(t, I) : new Unbound(x);
    case Const c: 
      return new Success(constType(c), I);
    case Lambda lam:
      Result r = 
        earlycheck(I, lam.declEffect(), /$\Gamma$/.with(lam.var(), lam.argty()), lam.body());
      if (!r.isSuccess()) return r;
      Success s = (Success) r;
      return new Success(new FunType(lam.argty(), lam.declEffect(), s.type), I);
    case App app:
      Result resa = earlycheck(/$\chi_0$/, /$\chi_m$/, /$\Gamma$/, app.fun);
      if (!resa.isSuccess()) return resa;
      Success sa = (Success) resa;
      /* We know sa.effect\($\chi_0$\$\chi_m$) is defined */
      if (!sa.type.isFun()) return new BadType(app.fun);
      Result resb =  earlycheck(/$\chi_0\rhd$/sa.effect, /$\chi_m$/, /$\Gamma$/, app.arg);
      if (!resb.isSuccess()) return resb;
      Success sb = (Success) resb;
      /* sb.effect\(sa.effect\($\chi_0$\$\chi_m$)) defined */
      if (!sa.type.arg.equals(sb.type)) return new BadType(app.arg);
      boolean goodEffect = hasResidual(sa.type.asFunc().latentEffect(),
                                       (sb.effect \ (sa.effect \ (/$\chi_0$/ \ /$\chi_m$/)));
      if (goodEffect) { 
        return new Success(sa.type.result, 
                    sa.effect /$\rhd$/ sb.effect /$\rhd$/ sa.type.asFunc().latentEffect()); 
      } else {
        return new BadEffect(app);
      }
    case If i: ...
    case While w: ...
  }
}
\end{minted}
    \caption{Algorithm for early effect errors.}
    \label{fig:algo}
\end{figure}

The key insight of the residual checking is apparent in the case for applications.
This case first type-checks the function position. If that subterm is already problematic then the underlying error is returned. Otherwise, the algorithm continues to ensure the function subterm has a function type, then type-checks the argument position \emph{with a modified progress effect} reflecting that it reduces after the function position (following \textsc{T-App}). If both subterms typecheck (in appropriate contexts), and the argument position is of an appropriate type for the function, then the algorithm checks that the residual corresponding to the final check in \textsc{T-App} is defined, returning success in that case.
Note that while there is only one residual check in the application case here, errors are still reported as early as possible in program order.  In particular:
\begin{itemize}
    \item If the function subterm's effect could not be sequenced after $\chi_0$, or the result would not leave the appropriate residual defined, then type-checking of that \emph{subterm} would fail.
    \item Similarly, if the argument subterm's effect could not be sequenced after $\chi_0$ and the function subterm's effect, type-checking of the argument would fail (note that the function's effect is passed into type-checking for the argument).
    \item The only residual not guaranteed to be defined is the final step of the residual checking, involving the function's \emph{latent} effect, whose interaction with the evaluation context is not implied by the function and argument subterm effects.
\end{itemize}
The cases for conditionals and loops are similar, checking subexpressions in program order, with only one new explicit residual check in each case.

\section{Implementation}
\label{sec:implementation}
We have implemented this approach in a prototype extension of the Checker Framework~\cite{PapiACPE2008,DietlDEMS2011} to 
support sequential effect systems.  
Currently we support the fragment of Java corresponding to a core object-oriented language: classes, methods (including checking a 
method override's effect is less than the original's), conditionals, while loops, calls, and switch statements. 
Exceptions are supported through a variant of Gordon's work on tagged delimited continuations~\cite{ecoop20b},
described in more detail in Appendix \ref{apdx:nonlocal}.

\begin{figure}[t]
    \begin{minted}[fontsize=\small]{java}
public abstract class EffectQuantale<Q> {
  public abstract Q LUB(Q l, Q r);
  public abstract Q seq(Q l, Q r);
  public abstract Q unit();
  public abstract Q iter(Q x);
  public abstract Q residual(Q sofar, Q target);
  public boolean LE(Q left, Q right) {
    return LUB(left, right).equals(right);
  }
  public boolean isCommutative() { return false; }
  public abstract ArrayList<Class<? extends Annotation>> getValidEffects();
}
    \end{minted}
    \caption{Framework interface to effect quantales with effects represented by type \mintinline{java}{Q}.}
    \label{fig:eqjava}
\end{figure}

The framework extension is parameterized by a choice of effect quantale, represented by the abstract class in 
Figure \ref{fig:eqjava}, which is parameterized by the representation type \mintinline{java}{Q} for a given system's 
effects.
It contains operations for $\sqcup$ (\mintinline{java}{LUB}), $\rhd$ (\mintinline{java}{seq}), unit, 
$-^*$ (\mintinline{java}{iter}), and residual checks (\mintinline{java}{residual}). 
Partiality is modeled by returning \mintinline{java}{null}. A default implementation of $\sqsubseteq$ 
(\mintinline{java}{LE}) is provided but can be overridden with more efficient implementations. 
\mintinline{java}{getValidEffects()} produces a list of Java annotation types the framework should recognize as being 
part of this effect quantale.

\mintinline{java}{isCommutative} indicates whether the effect quantale is commutative, which the framework uses to 
recover exhaustive error checking for such systems. In general, when the effect of an expression is 
$\alpha\rhd\beta\rhd\gamma$ and the residual $(\alpha\rhd\beta)\setminus \delta$ is undefined, the framework stops
 issuing errors about the rest of execution on the same path through the current method, because all residuals with more 
complete body effects (e.g., $(\alpha\rhd\beta\rhd\gamma)\setminus \delta$) will also be undefined, but not necessarily 
because of problems with the extensions (i.e., $\gamma$ may be fine if the problem is $\beta$, in which case further 
errors would be redundant). However, in the case of a commutative $\rhd$, the framework can exploit commutativity to 
give additional error messages. For example, if $(\alpha\rhd\beta)\setminus \delta$ is undefined as before, 
then $(\alpha\rhd\beta\rhd\gamma)\setminus \delta$ will also be undefined, but 
$\alpha\rhd\beta\rhd\gamma=\alpha\rhd\gamma\rhd\beta$, and it is possible that $(\alpha\rhd\gamma)\setminus \delta$ may 
be defined or undefined, independent of the residual involving $\beta$. 
This is precisely why the standard approach for 
join-semilattice effect systems of checking individual subexpressions' effects against a bound works and gives all 
appropriate errors: in our setting, the residual is defined as the bound itself as long as the effect so far is less 
than the bound, which via commutativity extends to the join of any non-empty subset of body effects being less than the 
bound.
Note however that our approach is general to any commutative effect quantale, including systems like must-effect analysis~\cite{mycroft16},
not just join semilattices.

General checking logic in the Checker Framework (as in the rest of the Java compiler) uses a visitor to traverse ASTs, 
rather than recursive traversals in Figure \ref{fig:algo}. The Java compiler provides the type environment as ambient 
state in this setting, and the implementation also maintains the current $\chi_m$ and $\chi_0$  as visitor
 state --- $\chi_0$ is maintained as a stack of subexpression effects which can be rewound to consider alternate paths 
(e.g., different branches of a conditional).
But the core algorithm is as demonstrated in the application case of Figure \ref{fig:algo}, with a single explicit 
residual check in each case.

We do make two kinds of extensions to the logic.
First, in addition to other varieties of loops (which are straightforward adaptations of \textsc{T-While}),
we handle additional language constructs present in Java: our handling of exceptions, breaks, and early returns
follows an extension of Gordon's work on effect systems for tagged delimited continuations~\cite{ecoop20b}.
Gordon defines a transformation of an arbitrary effect quantale that is ignorant of non-local control flow
to one that works with tagged delimited continuations. From this, typing rules for checked exceptions,
break statements, and early returns can be derived, which we implement.
We also extend this construction with residuals, and it is this extension which requires our shift to weak residuals:
intuitively, Gordon's construction tracks sets of possible effects for each execution path (normal execution,
for each thrown exception, etc.). In general, the \emph{greatest} possible residual for this construction may require
modeling an infinite set, while our implementation relies on finite sets.
Further details of our handling and residual are given in Appendix \ref{apdx:nonlocal}.

Second, when a type error is encountered, the algorithm does not immediately stop. It will immediately report the 
        error, but then sets a flag indicating an error has been found on the current path, suppressing further error 
        reporting. Then traversal continues in order to visit subexpressions that correspond to checking different 
        method bodies (i.e., lambda expressions and anonymous inner classes) and report errors there, which are 
        independent of errors in the surrounding code. When checking conditionals, if an error is encountered in one 
        branch the algorithm resets the flag for an error on the current path and checks the other branch as 
        if it were the only branch (including visiting code later in program order than the 
        whole conditional construct). This permits the implementation to report additional independent errors, rather 
        than allowing errors in one branch to shadow errors in the other.

Our extension inherits the Checker Framework's existing robust support for 
subtyping, and type generics.  While not used in any of our evaluations, the base implementation of the effect visitor 
extends the base implementation used for the Checker Framework's focus on type qualifiers~\cite{Foster1999}, so effect 
systems can in principle make use of type qualifiers to determine effects.

Effects are declared as Java annotations targeting method nodes.  This unfortunately requires a bit of boilerplate: each effect requires 10 lines of code, but 8 are identical across all declarations (import statements and meta-annotations for the Java compiler to allow them on method declarations and persist them in bytecode), with the remaining lines being the package declaration and one line for naming the actual annotation.

\paragraph{Performance}
The execution time of these checkers is dependent primarily on two factors.
The first is the underlying effect quantale: if computing sequencing, joins, iteration, and residuals for the underlying effect quantale is particularly slow, 
this will slow the whole framework.
The atomicity and reentrancy effect systems we have implemented (Section \ref{sec:eval}) both have very fast basic effect quantale operations. 
The second is the cost of working with the control effect transformation of the underlying effect system to handle exceptions, breaks, and non-local returns without 
individual effect systems needing to address them (Appendix \ref{apdx:nonlocal}).
In the common case (no non-local control flow) an effect represented as an object with an underlying effect representation, and two null pointers, so the operations 
have very little additional cost over ignoring non-local control entirely.
In code that contains non-local control flow or calls methods with checked exceptions, a prefix effect characterizing behavior up to each
break, non-local return, throw, and checked exception mentioned in the signatures of called methods. 
Composing such effects may trigger at most $n$ additional calls to the underlying effect quantale when each effect is tracking at most $n$ non-local behaviors, so the costs
do not grow significantly with code complexity.

Critically, since we analyze a single method at a time, even when those sets become large their performance impact is confined to the current method being checked. 
The only way for complexity of one method to influence the cost of checking another is via method annotations that expose latent effects for a 
variety of different thrown exceptions in addition to the non-exceptional method body effect. In general the number of such annotations that must appear is 
dependent not only on the program being analyzed, but also on the specific effect system in use.

In practice, the performance of our implementation, for the inexpensive effect quantales described in the next section, is on par with other existing pluggable type systems
in the Checker Framework.

\section{Evaluation}
\label{sec:eval}
The hypothesis underlying this work is that residuals offer a clear-cut way to localize sequential effect system 
errors to the earliest program location in program order where a mistake can be recognized, and that this 
useful precision for developers.

We have implemented two sequential effect systems in this framework to evaluate whether the error locations reported are accurate,
which we approach from two angles.
First, we reproduce part of Flanagan and Qadeer's evaluation of their atomicity effect system~\cite{flanagan2003atomicity}, where they
found atomicity errors in the then-current JDK (which have since been fixed). The original evaluation simply described the errors
as being found as a result of their analysis, with the implication that their tool used global reporting and thus all location of the error
was manually driven.
Second, we evaluate the accuracy of error reporting for the reentrancy effect system applied to non-reentrant database transactions.
This is a common situation across Spring Hibernate, JDBC, and other database systems, and is a situation with non-trivial interactions with
exceptions.
We find that the residual-based error locations are both predictable and accurate.
We evaluate these two systems because they cover both total and partial effect quantales, and because they are simple enough (both are finite
with 5 effects each) that we believe we can evaluate the residual-based locations of errors without becoming entangled
in deeper questions of how error messages are presented, which we believe is important future work for systems like trace effects via regular
languages (Section \ref{sec:formal_lang}) where the effects themselves, and consequently the relationship between the residual's existence
and the effects a programmer specifies, are more complex.

\subsection{Reproducing Atomicity Errors}
We have implemented the earlier of Flanagan and Qadeer's systems for static checking of atomicity via effects~\cite{flanagan2003tldi}
(introduced in Section \ref{sec:atomicity_bg}), 
in 201LOC, 50 of which were for the 5 effect declarations. 
A full reimplementation would require integration with a data race freedom system~\cite{toplas21} (since data races are 
non-atomic, while well-synchronized memory accesses are both-movers); this version assumes data race freedom.
We have run our atomicity checker with residual-based error reporting on several of the JDK1.4 Java classes
reported in Flanagan and Qadeer's evaluation to have errors, to check how accurate or useful the reported location is.

The most prominent example in their evaluation was an atomicity violation in the \lstinline|StringBuffer|
code in Figure \ref{fig:atomicity_bug}. The bug in the code is that while the code synchronizes on (locks)
the receiver \lstinline|this|, it performs \emph{two} atomic actions in the body (calls two atomic methods on
the argument, which is not locked), even though it is
supposed to be atomic. Our prototype reports the second atomic operation, which is the first subexpression in
the body that makes it impossible for the remainder of the method to have an overall atomic effect.

\begin{figure}[t!]
\begin{minted}{java}
@Atomic
@ThrownEffect(exception = StringIndexOutOfBoundsException.class, behavior = Atomic.class)
public synchronized StringBuffer append(StringBuffer sb) {
  if (sb == null) {
      sb = NULL;
  }
  int len = sb.length();
  int newcount = count + len;
  if (newcount > value.length)
    expandCapacity(newcount);
  sb.getChars(0, len, value, count); // <-- Error reported here
  count = newcount;
  return this;
}
\end{minted}
\caption{Excerpt from JDK1.4 \lstinline|StringBuffer| implementation.}
\label{fig:atomicity_bug}
\end{figure}

Flanagan and Qadeer report a similar bug in \lstinline|java.lang.String| (of JDK 1.4), shown in Figure \ref{fig:atomicity_bug2}.
There the same \lstinline|StringBuffer| methods are involved. Again our technique indicates the exact
point in the method beyond which an overall method body effect consistent with the annotation is impossible.
In this case, it saves the developer the trouble of looking at most of the method code.

\begin{figure}[t]
\begin{minted}{java}
@Atomic
public boolean contentEquals(StringBuffer sb) {
    if (count != sb.length())
        return false;
    char v1[] = value;
    char v2[] = sb.getValue(); // <-- Error reported here
    int i = offset;
    int j = 0;
    int n = count;
    while (n-- != 0) {
        if (v1[i++] != v2[j++])
            return false;
    }
    return true;
}
\end{minted}
\caption{Excerpt from JDK1.4 \lstinline|String| implementation.}
\label{fig:atomicity_bug2}
\end{figure}

Flanagan and Qadeer also analyzed other parts of the JDK 1.4, but we have had difficulty getting other classes from their evaluation to be accepted
with only minor modifications by the modern Java compiler the Checker Framework is a plugin to.

\subsection{Reentrancy}
We have also implemented Tate's system for reentrancy checking~\cite{tate13} (introduced in Section \ref{sec:reentrancy_bg}).
In this system, the \textsf{critical} effect describes code which is safe to use inside a critical section,
but which may not begin another (nested) critical section or end the current (presumed) critical section.
For our evaluation, rather than focusing on non-reentrant locks (which are little-used in Java),
we have focused instead on database transactions, as some database systems (notably Hibernate) do not support
nested transactions at all, while general-purpose database interfaces like JDBC leave the behavior of nested transactions up to the
particular backend chosen (making the use of nested transactions clearly wrong for some backends, and more generally
undefined behavior).

For our case study we focus on a variant of JBoss Hibernate's transaction API.\footnote{This is a variant of
the current API with checked exceptions. We must currently use a variant because the Checker Framework's support for stub
files (a means to externally annotate compiled JAR files) does not propagate our \mintinline{java}{@ThrownEffect}
annotation to the checker because it does not satisfy some in-built assumptions about which checker ``owns''
the annotation. While this certainly affects the real-world applicability of our checker as a tool, it does not
impact our evaluation of error reporting accuracy against an extracted copy of the API, and our focus is evaluation of
the residual-based error reporting.
}

\begin{figure}[t]
\begin{minted}{java}
@Entrant
@ThrownEffect(exception = TxException.class, behavior=Entrant.class)
public void docExample(SessionFactory factory) throws TxException {
  Session sess = factory.openSession();
  Transaction tx = null;
  try {
    tx = sess.beginTransaction();
  } catch (TxException e) {
    sess.close();
    throw e;
  }
  try {
    doWork(tx);
    tx.commit();
  } catch (TxException e) {
    tx.rollback();
    throw e;
  } finally {
    sess.close();
  }
}
\end{minted}
\caption{Slight refactoring of example from Hibernate documentation.}
\label{fig:hibernate}
\end{figure}

Consider the example code in Figure \ref{fig:hibernate} from Hibernate's documentation.\footnote{\url{https://docs.jboss.org/hibernate/orm/3.2/api/org/hibernate/Session.html\#beginTransaction()}}
Despite its small size, the main method \mintinline{java}{docExample} (correctly) handles several significant subtleties.
The main transaction itself is in a \mintinline{java}{try-catch} block,
as the session and transaction methods may throw an exception.
The \mintinline{java}{doWork()} method, a factored out transaction body,
 is marked \mintinline{java}{@Critical}.
The commit method is annotated as
\begin{minted}{java}
@Unlocking
@ThrownEffect(exception=TxException.class, behavior=Basic.class)
public void commit();
\end{minted}
indicating that if it succeeds it behaves as if unlocking (i.e., finishing the transaction),
while if it throws a \mintinline{java}{SQLException} the transaction remains open.
This ensures that if the body throws an exception, the catch block is checked assuming the code has not
yet finished the transaction, requiring the rollback attempt. Because rollback is typically
a last-resort fallback, it is annotated as
\begin{minted}{java}
@Unlocking
@ThrownEffect(exception=TxException.class, behavior=Unlocking.class)
public void rollback();
\end{minted}
since if it fails, the connection is almost certainly broken, and the database will automatically rollback the transaction after a timeout.

Unlike the atomicity system, some sequential compositions in the reentrancy effect system are already undefined (recall Figure \ref{fig:crit-effects}),
and would be immediately and locally rejected even without residual-based error detection, which is focused on cases where
composition is defined but has already committed the program being analyzed to a course already known to be incompatible with its
top-level effect specification. So, for example, attempting to start an additional nested transaction would have effect 
\mintinline{java}{@Locking}$\rhd$\mintinline{java}{@Locking},
which is undefined and would be rejected without our extensions. The additional cases which are rejected earlier due to 
residual-based error checking are 
related to the rows of Figure \ref{fig:crit-effects}'s $\rhd$ table lacking certain operations:
notice that each non-unit row of the table has either \textsf{locking} and \textsf{entrant} results, or \textsf{unlocking} and \textsf{critical}
results. The former rows (for $\mathsf{locking}\rhd-$ and $\mathsf{entrant}\rhd-$) correspond to cases where
the code has already committed to being code that must start running \emph{not} inside a transaction, so have no residual with
respect to a method-level annotation that the method's code should be able to start inside a transaction (i.e., have the \textsf{critical} or \textsf{unlocking})
effect.

The additional errors thus manifest in refactoring of the example code's actual
work into \mintinline{java}{doWork}. 
In large projects, it is easy to lose track of the intended execution context of a method~\cite{ecoop13},
sometimes resulting in developers incorrectly assuming they may need to construct some of that context themselves.
While directly starting a nested transaction
in the same syntactic scope as another transaction start would be undefined and therefore reported
precisely even without our extension (one of the original, informal arguments in favor of effect quantales being partial~\cite{toplas21}),
 factoring the body of the transaction out into this helper
method requires annotating the transaction body method with an effect. 
If it were annotated with \mintinline{java}{@Entrant},
the call site (specifically) would be rejected as undefined (since \mintinline{java}{@Locking}$\rhd$\mintinline{java}{@Entrant} is undefined).
Annotating it as \mintinline{java}{@Critical} (as we do) the call site is accepted, but the factored-out method would be 
rejected, raising the question
of where the error would be reported. Our technique reports the start of the nested transaction as the error location even if it is the first
line of code:
\begin{minted}{java}
@Critical
@ThrownEffect(exception=TxException.class, behavior=Critical.class)
public void doWork(Transaction tx) throws TxException {
  ...
  tx.begin(); // <-- error reported
  ...
}
\end{minted}
Thus also with an effect quantale with partially-defined compositions, residual-based error checking yields additional
precise error locations.

\section{Related Work}
The most closely-related work to ours is that on implementations of sequential effect systems.  The implementations we know 
of for concrete sequential effect systems~\cite{skalka2008types,flanagan2003atomicity} do not have error handling described 
in the corresponding publications, but are formalized in the standard way which corresponds to the all-at-once method 
behavior check.  It is possible that these systems implemented some kind of eager error reporting in the tools themselves, 
but the sources are no longer available and in any case these would be optimizations for specific effect systems. Our 
results establish a profitable eager error reporting strategy for \emph{many} sequential effect systems expressible as 
effect quantales (it is not necessarily the case that all effect quantales have a residual operator as we propose, but all 
those described as effect quantales in the literature~\cite{toplas21} have residual operators).

There have also been a number of generalized implementations of sequential effect system frameworks. 
Hicks et al.~\cite{hicks2014polymonadic} implement a general elaboration to polymonadic effects, which are equivalent to 
Tate's productors~\cite{tate13}. They use a constraint-based approach to determine which specific monad each expression
should be in (and therefore the effect of each expression).
Orchard and Petricek~\cite{orchard2014embedding} and Bracker and Nilsson~\cite{bracker2015polymonadic_haskell} implement an 
embedding of graded monads into Haskell, using typeclass constraints to define the composition and lifting operations. 
Because this is constraint-based, effect errors will be issued at an arbitrary program point corresponding to a failed 
constraint, which as in general type inference may be not directly related to the actual mistake in the program.
Bracker and Nilsson~\cite{bracker2016supermonads,bracker2018supermonads} later defined \emph{supermonads}, which 
generalize many monadic computation types by generalizing to an arbitrary number of parameters to a monadic type 
(vs. 1 for indexed monads~\cite{wadler2003marriage}, 2 for parameterized monads~\cite{atkey2009parameterised}), 
and can be used to express polymonads. They also added specialized support for supermonad constraints to Haskell's type inference.
Ultimately, because the constraints are still solved in an 
arbitrary solver-selected order, this suffers from the same problems with unpredictable error locations that exist in the 
normal Haskell implementations, where errors may be issued in unproblematic program locations, and program changes 
unrelated to the error may change where an error is reported by affecting constraint solving order.
Our approach yields predictable error locations with some guaranteed relevance to actual program errors.

More broadly, there is a wealth of work on better localizing type errors in constraint-based type inference. 
Many techniques have been applied with a wide variety of trade-offs.  
Most of this involves searching for a minimal program or type repair that results in inference 
succeeding~\cite{zhang2014toward,pavlinovic2015practical,pavlinovic2014finding,loncaric2016practical,chen2018systematic}, 
and reports the location whose term or type was assumed to change or whose type constraint was removed as the most likely 
error location (in general a program with a general type error may have many incompatible constraints,
so the smallest number of changes or removals that fixes the most incompatibilities is likely a source).
These approaches are all quite sensible, though both their approaches and setting differs significantly from ours.
None of this work on localizing type inference errors treats effects (notably, none of it targets a language in which 
monads are used to encode effects, though in principle these techniques could be applied to Haskell
and thus the Haskell embeddings of effects above). The assumptions available to us for our work are also much stronger 
in some ways than what is available to the general type inference localization problem. Because sequential effects are so 
closely coupled with evaluation order, there is a semantically-meaningful notion of best error location, while in general 
type inference the earliest inconsistency in program order may not be meaningfully related to the actual error location 
(hence the common practice there of exploring formalizations of the intuitive notions of ``minimum changes to fix'').
In addition, our effects have far more structure than typical type inference problems, because compared to general bags 
of constraints as in HM(X)~\cite{odersky1999type} or extensions thereof for 
object-orientation~\cite{pottier-icfp-98,TrifonovS96}, effect quantales afford a convenient algebraic 
characterization of an operation useful for error location: the residual.
As a result, our work is the first which takes an \emph{algebraic} approach to localizing effect errors.

\section{Conclusions}
We have proposed the first algebraic approach to localizing errors in sequential (flow-sensitive) effect systems,
by  exploiting the notion of a partial residual. 
This approach is guaranteed to give more precise error locations than the method-global (and therefore
highly imprecise) or constraint-based (and therefore unpredictable) techniques used in all prior sequential effect system
implementations, locations which are moreover guaranteed to have relevance to the actual program mistake.
We have implemented our technique for Java in a fork of the open source Checker Framework,
and shown that our technique gives specific meaningful error messages for previously studied bugs.

\bibliographystyle{splncs04}
\bibliography{effects,csg}

\appendix
\section{Further Residuated Effect Quantales}
\label{apdx:more}

\subsection{Parameterized Monads}
Parameterized monads~\cite{atkey2009parameterised} are monads indexed by models of state before and after execution of an expression, which Atkey showed can be used to capture session types~\cite{Honda2008}, answer type modification for delimited continuations~\cite{danvy1989functional}, and other examples; more broadly, they capture program logics as effects~\cite{Orchard_2020}.
Parameterized monads always have all residuals of effects with matching preconditions: 

$$(x,y)\setminus(x,z)=(y,z)$$

Interpreting the effects a pre- and post-state pairs as originally intended, there is of course no guarantee that any program exists with the residual as an effect (taking the parameterized monad capturing Hoare logic, there should be no program with effect $(\mathsf{true},\mathsf{false})$). 
In practice this is can in the worst case push error reports to the full method level,
as alluded to in the earlier discussion that residuals only assist with error localization when
residuals are sometimes undefined.
Notably, if a program fragment (roughly, prefix of a method's code) leaves effect $(x,z)$ ``left'' to execute, it can always
be followed by any program fragment with precondition $x$, and the residual will still be defined. 
In general, this means errors in a parameterized monad (as an effect quantale) will always appear either 
(1) when one computation fails to establish the precondition of the next in program order (including cases where no 
program fragment with that precondition exists), or (2) when effect completion is checked at the end of an execution path.  
These are the same points of effect-checking failure present in any implementation of parameterized monads,
but because our type-checking uses deterministic computation simulating program order, rather than using possibly-non-deterministic constraint solving that ignores program order, we will always observe the earliest error in program order.

\subsection{2-operation Commutative Effects}
Mycroft et al.~\cite{mycroft16} offer a 2-operation commutative effect system for \emph{must-do} analysis: where an effect describes a set of operations code is \emph{guaranteed} to do (modulo termination). Gordon~\cite{toplas21} models this as an effect quantale:
\begin{definition}[Must Effects]
    For a set $\Upsilon$ of events of interest:\\
    \eqdecl{\mathcal{P}(\Upsilon)}{X}{Y}{X\cup Y}{X\cap Y}{\emptyset}
\end{definition}
Note that the ordering on these effects is the \emph{opposite}  of the typical {inclusion} ordering for powersets: for two branches with effects (definite actions) $X$ and $Y$, the set of definite actions for a conditional uses the \emph{intersection} of their definite actions (those definitely occurring on both branches). In this case, the residual is:

\[ X\setminus Y = \{ y \in Y \mid y \not\in X\} \]
The right hand side above could be written simply as set difference, but we have reserved the standard notaton for set difference in this paper for residuals, in keeping with the literature on ordered semigroups rather than combinatorics.

\section{Extensions for Nonlocal Control Flow}
\label{sec:extensions}
\label{apdx:nonlocal}
There are two primary extensions to the formal system of Figure \ref{fig:typing} necessary for Java.

\paragraph{Exceptions}

Exceptions are implemented conceptually as a restriction of Gordon's work on sequential effects for tagged delimited continuations~\cite{ecoop20b}. The core idea is that each expression has not just one effect, but instead a combination of:
\begin{itemize}
    \item One effect bounding behaviors on all normal return paths (e.g., via a \mintinline{java}{return} statement or finishing execution of a \mintinline{java}{void} method), which may be absent for expressions which always throw (or later, break). We call this the \emph{underlying} effect.
    \item Effects for each checked exception the method may throw, capturing the behavior up to the time the exception was thrown.  We call this behavior up to the throw the \emph{prefix effect} of the throw.
\end{itemize}

Gordon~\cite{ecoop20b} describes a construction $\mathcal{C}(-)$ which transforms an effect quantale with no treatment of special behaviors into one with support for tagged delimited continuations with abort (as in Racket~\cite{Flatt2007}), and uses typing rules for those constructs to derive typing rules for other constructs such as exceptions and generators.  We give here a variant of a subset of this construction based on the ideas above, sufficient for treating Java's checked exceptions.

\begin{definition}[Exception Effect Quantale]
    Given an effect quantale $Q$, and poset of set of checked exceptions $X$ define the \emph{exception effect quantale} $\mathcal{E}_X(Q)$ to be the effect quantale with carrier $(\mathsf{option}~Q)\times \mathsf{set}~(E\times Q)$ (a pair of optional \emph{underlying} effect and a set of \emph{control effects}):
    \begin{itemize}
        \item $I=(I_Q,\emptyset)$
        \item $(\chi,X)\rhd(\chi',X')=\left\{\begin{array}{ll}
            (\mathsf{None},X) & \textrm{when $\chi=\mathsf{None}$}\\
            (\mathsf{None}, X\cup(\chi\rhd X')) & \textrm{when $\chi'=\mathsf{None}$}\\
            (\chi\rhd_Q \chi', X\cup(\chi\rhd X')) & \textrm{otherwise}\\
            \end{array}\right\}$
        \item $(\chi,X)\sqcup(\chi',X') = (\chi\sqcup_Q \chi', X\cup X')$
    \end{itemize}
Above, the notation $\chi\rhd x$ for $\chi$ an element of the (underlying) effect quantale $Q$ and $x$ a set of control effects as above is defined as:
\[
    \chi\rhd x = \mathsf{map}\;(\lambda (\chi',x)\ldotp (\chi\rhd_Q\chi,x))
\]
That is, it extends the prefix effect on the left with the non-throwing behaviors that preceeded. If any resulting use of the underlying effect quantale's operators is undefined, so is the corresponding operator on the exception effect quantale.
\end{definition}

Intuitively, the rule for try blocks matches the prefix effect of an exception with the behavior of the corresponding 
catch block, and non-exceptional runs of the catch block, following that prefix, are joined (via $\sqcup$) with
the normal effect of non-exceptional executions of a try block.

, but first we describe the residual operator on this construction, 
Assuming a residual operator on the underlying effect quantale, we can describe a \emph{weak} residual operator
on $\mathcal{E}_X(Q)$:
\[
    (\mathsf{Some}(\chi),X)\setminus(\mathsf{Some}(\chi'),X') = (Some(\chi\setminus{}\chi'), \mathsf{excResiduals}(\chi,X'))
\]

when the underlying residual is defined and
\[\forall (\chi_x,x)\in X\ldotp \exists \chi_{x'},x'\ldotp \chi_x\sqsubseteq\chi_{x'}\land x\le x'\]
where
\[
    \mathsf{excResiduals}(\chi,X) = \mathsf{map}~(\lambda(\chi_x,x)\ldotp (\chi\setminus\chi_x,x))~(\mathsf{filter}~(\lambda(\chi_x,x)\ldotp \mathsf{defined}(\chi\setminus\chi_x))~X)
\]
i.e., \textsf{excResiduals} filters exception behaviors which have a residual with the behavior so far (i.e., those for which the underlying behavior so far is a prefix of the permitted exception prefix), and then replaces those behaviors with that underlying residual (since the overall residual wants behaviors that can be sequenced after the behavior so far and still be less than the target behaviors).
This construction is the reason we must work with \emph{weak} residuals: depending on the underlying effect quantale
transformed in this way, the \emph{greatest} possible quotient result may be an infinite set which we cannot
represent programmatically; our implementation works only with finite sets.

Visitor state includes not only the aforementioned markable stack of effects, but also a map from exception types to effects. When a throw is visited, the thrown expression is first recursively visited, then the exception map is updated with the current base effect ($\chi_0$) and the regular return effect is marked impossible to indicate no non-exceptional return paths exist for the throws clause. (The impossible mark can be removed by visitors higher up the tree, such as when the throw is in one branch of a conditional.)

Method calls with checked exceptions are handled similarly, but instead state is updated for each checked exception that may be thrown, and the base effect is extended by the normal latent effect to accommodate standard return paths.

The description thus far corresponds precisely to a subset of Gordon's system~\cite{ecoop20b}. One complication of Java exceptions not treated by Gordon's system is the subtyping relationship between exceptions.  At the cost of some precision, exception subtyping is handled as follows:
\begin{itemize}
    \item If a possible throw of an exception type not already in the exception map is encountered, it is handled as above
    \item For a possible throw of an exception that is a \emph{supertype} of one or more exceptions already in the exception map, the supertype is added to the map with an effect that is the least upper bound of the current path \emph{and} the effects for all known subtypes of it that may be thrown.
    \item For a possible throw of an exception that is a \emph{subtype} of one or more exceptions already in the exception map, the subtype's own path is handled as above, but all known supertypes' exceptional effects are updated to the least upper bound of the known effect and the new prefix effect. 
\end{itemize}
This maintains the invariant that if the exceptional effect map has entries for both \mintinline{java}{Sub} and \mintinline{java}{Sup} where \mintinline{java}{Sub extends Sup}, the effect for \mintinline{java}{Sup} is always greater than or equal to the effect tracked for \mintinline{java}{Sub}. If the mentioned least-upper-bounds do not exist, an error is issued.

Try-catch blocks are given effects in line with Gordon's work~\cite{ecoop20b}, as the least upper bound of the try block's normal effect and any valid pairing of a tracked exceptional effect and a corresponding catch block (i.e., the least-upper-bound of the effects of every path through the try-catch). After the try-catch, any caught exceptions are removed from the map. At the method level, any remaining exceptions are compared against exceptional return effects, e.g.,
\begin{minted}[fontsize=\small]{java}
    @ExceptionEffect(IOException.class, Atomic.class)
\end{minted}
indicates that executions of the annotated method that finish by throwing an \mintinline{java}{IOException} have effect \mintinline{java}{@Atomic}.

One final shift is that residual checks are now performed not only with regards to the base (normal-return) effect, but also with regard to annotated exceptional behaviors. So errors are only issued if the current execution effect has no residual with (is not an effect prefix of) the method's base effect \emph{or} any of the annotated exception effects
(i.e., it is possible that $\chi\setminus\chi_m$ for effects tracking exceptions contains \emph{only} exception-throwing paths).

\paragraph{Early Returns and Breaks}
Early returns and breaks are handled similarly to exceptions, as an additional set of prefix-before-nonlocal-control
effects. At each source construct which may be the target of a break (loop boundaries, switch statements), break behaviours are flattened
into the underlying behavior. Likewise, early-return effects are flattened at method boundaries.
Both of these accumulate extended prefixes when they occur in loops, just as Gordon~\cite{ecoop20b} treats abort effects;
the identity of the AST node a break or early return would target is essentially used as a unique tag for
those non-local control transfers in Gordon's work.

\end{document}